\documentclass[12pt]{article}

\usepackage{amsmath}
\usepackage{cite}
\usepackage{graphicx}

\usepackage[unicode,linktocpage=false,plainpages=false,pdfpagelabels=false]{hyperref}

\newcommand{\m}{\mu}
\newcommand{\n}{\nu}
\newcommand{\ls}{\left(}
\newcommand{\rs}{\right)}
\newcommand{\al}{\alpha}
\newcommand{\be}{\beta}
\newcommand{\ff}{\varphi}
\newcommand{\te}{\theta}
\newcommand{\ga}{\gamma}
\newcommand{\sh}{\sinh}
\newcommand{\ch}{\cosh}

\begin{document}

\title{
Global Embedding of the Reissner-Nordstr\"om Metric\\
in the Flat Ambient Space
}
\author{S.A.~Paston\thanks{E-mail: pastonsergey@gmail.com},
A.A.~Sheykin\thanks{E-mail: a.sheykin@spbu.ru}\\
{\it Saint Petersburg State University, St.-Petersburg, Russia}
}
\date{\vskip 15mm}
\maketitle

\begin{abstract}
We study isometric embeddings of non-extremal Reissner--Nord\-str\"om metric describing a charged black hole. We obtain three new embeddings in the f\/lat ambient space with minimal possible dimension. These embeddings are global, i.e.\ corresponding surfaces are smooth at all values of radius, including horizons.
Each of the given embeddings covers one instance of the regions outside the horizon, one instance between the horizons and one instance inside the internal horizon.
The lines of time for these embeddings turn out to be more complicated than circles or hyperbolas.
The obtained embeddings are also smooth at all values of radius for extremal and hyperextremal black holes.
\end{abstract}

\newpage

\section{Introduction}\label{p1}

It is known that any $d$-dimensional Riemannian manifold can be locally isometrically embedded in an $N$-dimensional f\/lat ambient space, with $N=d(d+1)/2$. Then we can describe this manifold in terms of the embedding function $y^a (x^\mu)$, so the metric on the manifold becomes induced (see, for instance, \cite{goenner}):
\begin{gather}
      \partial_\mu y^a (x) \partial_\nu y^b (x) \eta_{ab}=g_{\mu\nu} (x),
      \label{1.1}
\end{gather}
where $\eta_{ab}$ is a f\/lat metric of the ambient space, $\m,\n=0,1,2,3$ and $a$, $b$ range over $N$ values.

If a manifold has any symmetries, the number of the ambient space dimensions can be smaller than $d(d+1)/2$. The dif\/ference between the dimension of the f\/lat ambient space and the dimension of the original manifold is called the embedding class, $p=N-d$. It is easy to see that $p\leq 6$ if $d=4$. In particular, we have $p=1$ for constant curvature spacetimes, while for spherically-symmetric spacetimes one obtains $p \leq 2$ \cite{schmutzer}. Since the embedding class is invariant, the embedding is useful for classif\/ication of exact solutions of the Einstein f\/ield equations~\cite{schmutzer}. Note that the exact form of the embedding is not required for determination of the embedding class, so  for this purpose one prefers the Gauss--Codazzi--Ricci equations~\cite{schmutzer}, for which at $p\le 2$ we have some regular methods of solving. When $p>2$ such methods do not  exist, and no systematic classif\/ication of manifolds with $p>2$  has still been performed.

The embedding can be of use for studying geometric properties of Riemannian manifolds, and in this case the explicit form of the embedding function is, of course, necessary.  A possibility of the embedding also allows one to formulate the gravity as a theory of a four-dimensional surface in a f\/lat ambient space \cite{regge, deser}, similar to the string theory formulation. The developments in this f\/ield include, for instance, the canonical formulation of this theory and the consideration of its quantization~\cite{davkar}, the canonical formulation with the imposing of additional constraints (proposed in~\cite{regge}), which provide the elimination of ``extra solutions'' \cite{statja18,statja24}, the reformulation in terms of a f\/ield theory in the f\/lat ambient space~\cite{statja25} and other works \cite{pavsic85let,tapia,maia89,bandos,statja26}. One can f\/ind a detailed bibliography of the embedding-related questions in~\cite{tapiaob}.

Since the system (\ref{1.1}) is a system of nonlinear PDE's, the construction of an embedding is a~dif\/f\/icult problem. In the general case there is no regular methods of solving (\ref{1.1}). But it greatly simplif\/ies if the manifold has a high enough symmetry. In particular, in case of a Friedmann, static spherical or plane-wave symmetry this system transforms into the system of ODE's and can be exactly solvable. Thus for manifolds with high symmetry a relatively large number of explicit embeddings was found; see, for instance, \cite{rosen65} and \cite{collinson68}.

However, even in this case the classif\/ication of the solutions of the system (\ref{1.1}) remains non\-tri\-vial. Fujitani et al.~\cite{fujitani} attempted to construct all the possible embeddings of the Schwarzchild metric, but their approach turned out to be not enough general. In \cite{statja27} the method based on the representation theory was proposed, which allows to enumerate all possible ways to  realize certain symmetries of the manifold, and to construct an explicit embedding on this basis. Using this method, all possible minimal (i.e., in the f\/lat ambient space with minimal dimension) embeddings of the Schwarzchild metric were found.

In the present work we use the same method in order to construct all global minimal embeddings of the Reissner--Nordstr\"om (RN) metric describing  a charged black hole with a mass $m$ and a charge $q$ (its sign can be $+$ or $-$, but hereafter all formulas will include only $|q|$ and we will write $q$ instead of $|q|$ for the sake of simplicity). The corresponding  line element has the form
\begin{gather}
      {\rm d}s^2=\left(1-\frac{2m}{r}+\frac{q^2}{r^2}\right){\rm d}t^2-\left(1-\frac{2m}{r}+\frac{q^2}{r^2}\right)^{-1}{\rm d}r^2
-r^2 \ls {\rm d}\te^2+\sin^2\te  {\rm d}\ff^2\rs
      \label{1}
\end{gather}
(we use the geometric system of units where  $m$ and $q$ has the dimension of length).
We restrict ourselves to consideration of  the most nontrivial case of the so-called ``non-extremal''  charged black hole, where $q<m$ and two horizons exist:
\begin{gather*}
      r_\pm = m\pm\sqrt{m^2-q^2}.
\end{gather*}
The consideration of other cases is similar. It is known that  RN metric can be minimally embedded in a six-dimensional f\/lat ambient space because the embedding class for it is $p=2$~\cite{Pandey}.

By the term ``global embedding''  we mean an embedding which corresponds to a 4-dimen\-sional surface which is smooth (analytical) for {\it all} $r>0$, including {\it both} horizons $r=r_\pm$. It should be especially emphasized because in some works the term GEMS (global embedding Minkowski space) is used for the embeddings which are smooth only for some value areas of $r$. As far as the authors know, at the present time there is no known global (in the above sense) embedding of the RN metric, even for  greater dimensions of the ambient space. In the work~\cite{hong2003} is discussed the continuation to the region with $r<r_-$ of the embedding  for the RN metric proposed in~\cite{deserlev99}, but the form of embedding function given there turns out to be complex in this region, and if one wants to make it real, it is necessary to change the signature of the ambient space in order to obtain a correspondence with the RN metric (for details see Section~\ref{p2}).

Note that besides the constructing of a full metric embedding which is considered in this work one can study an easier problem of construction of the embedding diagrams, when one looks for certain submanifolds of the Riemannian space. For the RN metric such studies were performed in \cite{gr-qc/0305102,gr-qc/0307056,gr-qc/0605104}.

The explicit form of the global minimal RN metric embeddings given in this work (as well as the global minimal Schwarzschild metric embeddings given in \cite{statja27}) can be of use for studying the relation between Hawking radiation and the
Unruh radiation detected by observer moving on the embedding surface in the ambient space.
In \cite{deserlev97,deserlev98,deserlev99} such a relation was discovered for various metrics with the horizons and for some embeddings of these metrics.
After this discovery the thermodynamic properties of black holes were investigated in the same way in many works, see for example \cite{hong2003,lemos, chen2005}.
However, preliminary calculations show that there is no mapping between Hawking ef\/fect and Unruh ef\/fect in ambient space for the new global minimal embeddings for RN metric and Schwarzschild metric. A detailed study of this question is beyond the scope of this paper.
Note also that there is a possibility to study the thermodynamical properties of the black holes using only the embedding diagrams mentioned above (instead of full embeddings) which are correspond to an embedding of two-dimensional $(t-r)$-submanifold into a f\/lat space. Such an approach was proposed in work \cite{banerjee10}.

In Section~\ref{p2} we list the previously known embeddings, classify  them by type of realization of the shift of the  parameter~$t$, and we discuss their properties. In Section~\ref{p3} we present the results of an application of the method proposed in~\cite{statja27} to the problem of constructing global RN metric embeddings in a six-dimensional f\/lat ambient space. Three new embeddings with the RN metric symmetry smooth for all $r>0$ are obtained.

\section{Known embeddings of the RN metric}\label{p2}

Although an exact local embedding of the Schwarzchild metric was found \cite{kasner3} just f\/ive years after founding the metric itself, an embedding for  the RN metric found in about the same time was constructed much later. To all appearance, the historically f\/irst embedding was published by Rosen~\cite{rosen65}. It is completely similar to the Kasner embedding for the Schwarzchild metric, and has the form
\begin{gather}
  y^0  = \sqrt {1 - \frac{2m}{r}+ \frac{q^2}{r^2}} \cos t,\nonumber\\
  y^1  = \sqrt {1 - \frac{2m}{r}+ \frac{q^2}{r^2}}  \sin t,\nonumber\\
  y^2  = r \cos\theta,\nonumber\\
  y^3  = r \sin\theta \cos\ff,\nonumber\\
  y^4  = r \sin\theta \sin\ff,\nonumber\\
  y^5  = \int {\rm d}r \sqrt{\frac{(m/r^2-q^2/r^3)^2+1}{1-2m/r+q^2/r^2}-1},
\label{r1}
\end{gather}
if $(++----)$ is the signature of the ambient space.
In this embedding the shifts of $t$ are realized as  ${\rm SO}(2)$-rotations in a two-dimensional subspace of the ambient space; such embeddings can be called ``elliptical''. As is easily seen, this embedding covers only the region outside the external horizon $r>r_+$ and cannot be continued under it.

In the article \cite{RosenRN} Rosen proposed one more embedding, now constructed similarly to the Fronsdal embedding~\cite{frons} for the Schwarzchild metric:
\begin{gather}
  y^0  = \sqrt {1 - \frac{2m}{r}+ \frac{q^2}{r^2}}  \sh t,\nonumber\\
  y^1  = \sqrt {1 - \frac{2m}{r}+ \frac{q^2}{r^2}}  \ch t,\nonumber\\
  y^2  = r \cos\theta,\nonumber\\
  y^3  = r \sin\theta \cos\ff,\nonumber\\
  y^4  = r \sin\theta \sin\ff,\nonumber\\
  y^5  = \int {\rm d}r \sqrt{\frac{1-(m/r^2-q^2/r^3)^2}{1-2m/r+q^2/r^2}-1},
\label{r2}
\end{gather}
with the signature $(+-----)$. Here the shifts of $t$ are realized as ${\rm SO}(1,1)$-rotations, and such embeddings can be called ``hyperbolic''. In the form~(\ref{r2}) this embedding covers only $r>r_+$ region, but it can be easily modif\/ied (following the idea of~\cite{frons}) to continue it under the exterior horizon. Unfortunately, it is still impossible to make it smooth for all $r>0$.

In \cite{Plazowski} an embedding in $(2{+}6)$-dimensional space was proposed with the same realization of shifts of $t$ as in~\cite{frons}. It is a continuation of an embedding with the same components of the embedding function~$y^{0-4}$ as in~(\ref{r2}), but with three functions of $r$ instead of~$y^5$. Formally such an embedding covers all values of~$r>0$, but at both horizons one of the embedding functions components becomes inf\/inite, so the manifold splits into three parts which are not connected between each other. The embedding \cite{Plazowski}, in contrast to the embeddings~(\ref{r1}),~(\ref{r2}) can be expressed in terms of elementary functions.

The f\/irst embedding that continues under an exterior horizon was proposed in~\cite{Ferraris}. It is an embedding in a $(3{+}6)$-dimensional space.
The form of this embedding is quite similar with that one proposed in~\cite{Plazowski}. It is smooth on the exterior horizon $r=r_+$, but on the interior horizon $r=r_-$ one component of the embedding function becomes inf\/inite. This embedding, as the previous one, can be expressed in terms of elementary functions.

A simpler RN metric embedding which continues under the exterior horizon was obtained in Deser and Levin work~\cite{deserlev99}:
\begin{gather}
\begin{array}{@{}ll}
r\ge r_+:        &  r_-<r\le r_+:\\
\displaystyle  y^0=\be^{-1}\sqrt{1-\frac{2m}{r}+\frac{q^2}{r^2}} \sh(\be t),\qquad &
\displaystyle y^0=\pm \be^{-1}\sqrt{\frac{2m}{r}-1-\frac{q^2}{r^2}} \ch(\be t),\vspace{1mm}\\
\displaystyle y^1=\pm \be^{-1}\sqrt{1-\frac{2m}{r}+\frac{q^2}{r^2}} \ch(\be t),  \qquad &
\displaystyle y^1=\be^{-1}\sqrt{\frac{2m}{r}-1-\frac{q^2}{r^2}} \sh(\be t),
\end{array}\nonumber\\
\hspace*{40mm}
\begin{array}{@{}l}
y^2=r\cos\te,\\
y^3=r\sin\te \cos\ff,\\
y^4=r\sin\te \sin\ff,\\
\displaystyle y^5=\int  {\rm d}r  \sqrt{\frac{r^2 (r_++r_-)+r^2_+(r+r_+)}{r^2 (r-r_-)}},\vspace{1mm}\\
\displaystyle y^6=\frac{2\sqrt{r_+^5 r_-}}{(r_+-r_-)r},
\end{array}\label{sp1}
\end{gather}
where $\be=(r_+-r_-)/(2r_+^2)$.

It is an embedding in a 7-dimensional space with the signature $(+-----+)$. It covers the region with $r>r_-$; in this region, similarly to  the  Fronsdal embedding of the Schwarzschild metric, it describes a maximal analytical extension of a Riemannian space: two copies of \mbox{$r>r_+$} region and two copies of $r_-<r<r_+$ region, corresponding to black and white holes (regions~I,~IV, II, III on the f\/igure given in Section~\ref{p3}).
 At  $r=r_+$ this embedding is smooth, though it cannot be smoothly continued in $r<r_-$ region.

In \cite{hong2003} the author attempted to use for $r<r_-$ the same form of an embedding function which is used for  $r>r_+$ in (\ref{sp1}), but it is  easy to see that in this case $y^5$ becomes imaginary. If one changes the sign under the radical in order to make $y^5$ real, the signature of the ambient space needs to be changed on $(+----++)$ in order to obtain the correspondence with the RN metric. Thus it is impossible to construct a unif\/ied embedding for all $r>0$ using this form of the embedding function; however, the fact that it smoothly describes the regions on both sides of the horizon $r=r_+$ allows to use it for studying of thermodynamic properties of black holes, see \cite{deserlev99}. A remarkable feature of this embedding is the allowance of $q\to 0$ limit, in which $y^6$ vanishes and the embedding (\ref{sp1}) becomes the Fronsdal embedding for the Schwarzschild metric.
Generalization of the embedding  (\ref{sp1}) to the case of a charged black hole in an arbitrary dimension in the presence of a nonzero cosmological constant was constructed in  \cite{lemos}.

\section{New global minimal embeddings of the RN metric}\label{p3}
As mentioned in the Introduction, the RN metric has the embedding class $p=2$, so an embedding in 6-dimensional space is minimal for it. From~(\ref{1}) we can see that the RN metric (as the Schwarzchild metric) has a symmetry ${\rm SO}(3)\times T^1$ where $T^1$ corresponds to the shifts of~$t$. Thus  in order to found the RN metric embeddings with the same symmetry one can use all types of the embedding functions constructed for the Schwarzchild metric in \cite{statja27} by the analysis of symmetry group representation.

If one is interested only in local embeddings which cover some regions of $r$, one can use all the six embedding types given in \cite{statja27}. However, not all of them are suitable for constructing the global embeddings which smoothly cover all values of $r>0$. The embeddings analogous to Kasner (i.e.\ elliptic, (\ref{r1}) is of this type) and to Fujitani--Ikeda--Matsumoto (this embedding is called sometimes ``parabolic'') must be dropped, since they cover only regions with a def\/inite sign of metric component~$g_{00}$, and in the most interesting case of the RN metric with $q<m$ this component changes its sign.

 It is also impossible to construct an embedding analogous to the Fronsdal one (i.e.\ hyperbolic, (\ref{r2}) is of this type) since such an embedding contains only one arbitrary parameter (the dimensional factor in the argument of hyperbolic functions which can be inserted in (\ref{r2})) by which one can ensure the smoothness at one of the horizons, but not at both simultaneously.

The three remaining  embedding types which were used in~\cite{statja27} for the Schwarzchild metric can be applied to the construction of global embeddings of the RN metric. In all these embeddings the shifts of $t$ are realized  under the form of certain rotations accompanied by shifts.

Although in this paper we restrict ourselves to the detailed consideration of "non-extremal" charged black hole ($q<m$), it must be noted that the embeddings obtained below remain smooth when $q\ge m$. It means that they are global embeddings for extremal RN black hole metric ($q=m$) as well as for hyperextremal ($q>m$).

\subsection{Spiral embedding}\label{p31}

Consider the form of the embedding function which was used in the construction of an asympto\-ti\-cal\-ly f\/lat embedding of the Schwarzchild metric (see~\cite{statja27}, in this case the shifts of $t$ are realized as ${\rm SO}(2)$-rotations with shifts, such an embedding may be classif\/ied as ``translation-elliptical''). If one substitutes it in the equation~(\ref{1.1}) with the RN metric in the right hand side, the resulting equation is satisf\/ied by the embedding function
\begin{gather}
  y^0 = f(r)\sin\ls\alpha t' + u_1 \ls\frac{2mr}{q^2}\rs\rs,\nonumber\\
  y^1 = f(r)\cos\ls\alpha t' + u_1 \ls\frac{2mr}{q^2}\rs\rs,\nonumber\\
  y^2 = \frac{\sqrt{b^2+m^2-q^2}}{q}  t',\nonumber\\
  y^3 = r \cos\theta,\nonumber\\
 y^4 = r \sin\theta \cos\ff,\nonumber\\
  y^5 = r \sin\theta \sin\ff\label{2}
\end{gather}
with the signature  $(++----)$.
Here $b>0$\footnote{In case of global embedding of hyperextremal black hole $b>\sqrt{q^2-m^2}$.},
\begin{gather}
\alpha = \frac{4m^3}{q^3\sqrt{m^2+b^2}},\qquad
f(r) =\frac{\sqrt{(mr-q^2)^2+b^2 r^2}}{\al q r},\qquad
t'=t+h_1 \ls\frac{2mr}{q^2}\rs,\label{sp2}
\end{gather}
and the functions $u_1(z)$, $h_1(z)$ are given by the integrals:
\begin{gather}\label{sp3}
u_1(z) = -\frac{4m^3}{\sqrt{m^2-q^2+b^2}} \int {\rm d}z \frac{\sqrt{P_1(z)}}{z^2(m^2(z-2)^2+b^2 z^2)},
\\
\label{sp4}
h_1(z) = \frac{4m^3}{\al\sqrt{m^2-q^2+b^2}} \int  {\rm d}z \frac{\sqrt{P_1(z)}}{z^2(q^2z^2-4m^2(z-1))},
\end{gather}
where
\begin{gather*}
P_1(z)=\frac{q^2z^2(z-2)^2}{4m^2}\!+\!\frac{(z-1)^2(m^2(z-2)^2(1+z+z^2+z^3)+b^2(4+z^2+z^3+z^4+z^5))}{m^2+b^2}.
\end{gather*}

It is easy to see that the function $P_1(z)$ under the radical in~(\ref{sp3}),~(\ref{sp4}) is strictly positive. This fact, coupling with a strict positiveness of the bracketed term in the denominator of the integral in~(\ref{sp3}), means the smoothness of~$u_1(z)$ for all $z>0$. As a consequence, the embedding function~(\ref{2}) describes the surface which is smooth for all $r>0$ since $f(r)$  turns to be smooth in this region. The surface points which correspond to $r\to0$ tend to inf\/inity since in this limit $f(r)\to\infty$ and the radius of a spiral corresponding to f\/ixed values of~$r$, $\te$, $\ff$ is given by~$f(r)$.

Thus the embedding (\ref{2}) is global (i.e.\ smooth for all $r>0$). We call it a ``spiral embedding''.
Note that when we describe the appearing smooth surface in terms of the coordinate~$t$
(just in this case the metric takes the form~(\ref{1})),
the coordinate singularity occurs, see Section~\ref{p34} for details.
The situation is similar for the embeddings listed in the next two sections.

\subsection{Exponential embedding}\label{p32}
Now let us use the form of the embedding function which used for construction of a Davidson--Paz embedding of the Schwarzchild metric (see~\cite{statja27}, in this case the shifts of $t$ are realized as~${\rm SO}(1,1)$ rotations with translations, this type is ``translation-hyperbolical''). Substituting it in the equation (\ref{1.1})  we obtain the solution
\begin{gather}
  y^0 = \ga t',\nonumber\\
  y^1 = \frac{1}{2\beta} \left(e^{-\be t' - u_2\ls 2mr/q^2\rs} -
 \ls 1-\frac{2m}{r}+\frac{q^2}{r^2}-\ga^2 \rs  e^{\beta t' + u_2\ls 2mr/q^2\rs}
 \right),\nonumber\\
  y^2 = \frac{1}{2\beta} \left(e^{-\be t' - u_2\ls 2mr/q^2\rs} +
 \ls 1-\frac{2m}{r}+\frac{q^2}{r^2}-\ga^2 \rs  e^{\beta t' + u_2\ls 2mr/q^2\rs}
 \right),\nonumber\\
  y^3 = r \cos\theta,\nonumber\\
  y^4 = r \sin\theta \cos\ff,\nonumber\\
  y^5 = r \sin\theta \sin\ff\label{sp6}
\end{gather}
with the signature  $(++----)$.
Here $\ga>1$,
\begin{gather}\label{sp7}
\beta=\frac{4m^3}{q^4 \sqrt{\gamma^2-1}},\qquad
t'=t+h_2 \ls\frac{2mr}{q^2}\rs,
\end{gather}
and the functions $u_2(z)$, $h_2(z)$ are given by the integrals:
\begin{gather}\label{sp8}
u_2(z) = \frac{m}{\ga q} \int  {\rm d}z \frac{2m q\ga z(z-2)+\sqrt{P_2(z)}}{z^2((\ga^2-1)q^2z^2+4m^2(z-1))},
\\
\label{sp9}
h_2(z) = \frac{m}{\be\ga q} \int  {\rm d}z \frac{\sqrt{P_2(z)}}{z^2(q^2z^2-4m^2(z-1))},
\end{gather}
where
\begin{gather}\label{sp10}
P_2(z)=4m^2q^2z^2(z-2)^2+4q^6\be^2z^4(z-1)^2+(2m)^4(z-1)^2\big(4+z^2+z^3+z^4+z^5\big).
\end{gather}

The function $P_2(z)$ under the radical in~(\ref{sp8}),~(\ref{sp9}) is strictly positive. The bracketed term in the denominator of the integral~(\ref{sp8}) vanishes in only one point, but we can prove that  in this point the corresponding numerator also vanishes, so $u_2(z)$ is smooth for all $z>0$.
From this we conclude that the embedding function~(\ref{sp6}) describes the surface which is smooth for all $r>0$. As in the spiral embedding, the surface points which corresponds to $r\to0$ tend to inf\/inity.

Thus the embedding (\ref{sp6}) is also global (i.e.\ smooth for all $r>0$). We call it the ``exponential embedding''. It should be noted that one can construct a global embedding like~(\ref{sp6}) with $\ga=1$, in this case the formulas~(\ref{sp7})--(\ref{sp10}) take another form, but in this case it is more dif\/f\/icult to determine the values of the parameters which ensure positiveness of the appearing expressions in the radicals.

\subsection{Cubic embedding}\label{p33}

Now let us use the form of the embedding function which is used for construction of a ``cubic in respect to time''  embedding of the Schwarzchild metric (see \cite{statja27}, in this case the shifts of $t$ are realized as null rotations with translations, this type is ``translation-parabolic''). Substituting it in the equation (\ref{1.1}) we obtain the solution
\begin{gather}
  y^0 = -\frac{q^4}{8m^3}\ls 1-\frac{2m}{r}+\frac{q^2}{r^2}\rs+\frac{2m^3}{q^4}t'^2,
\nonumber\\
  y^1 = u_3 \ls\frac{2mr}{q^2}\rs-\frac{1}{4}\ls 1-\frac{2m}{r}+\frac{q^2}{r^2}\rs t'+
\frac{4m^6}{3q^8}t'^3-t',
\nonumber\\
  y^2 = u_3 \ls\frac{2mr}{q^2}\rs-\frac{1}{4}\ls 1-\frac{2m}{r}+\frac{q^2}{r^2}\rs t'+
\frac{4m^6}{3q^8}t'^3+t',
\nonumber\\
  y^3 = r \cos\theta,\nonumber\\
  y^4 = r \sin\theta \cos\ff,\nonumber\\
  y^5 = r \sin\theta \sin\ff\label{sp11}
\end{gather}
with the signature  $(++----)$.
Here
\begin{gather}\label{sp12}
t'=t+h_3\!\ls\frac{2mr}{q^2}\rs,
\end{gather}
and the functions $u_3(z)$, $h_3(z)$ are given by the integrals:
\begin{gather}\label{sp13}
u_3(z) = \frac{q}{4m} \int  {\rm d}z \frac{\sqrt{P_3(z)}}{z^4},
\\
\label{sp14}
h_3(z) = \frac{q^3}{2m} \int  {\rm d}z \frac{\sqrt{P_3(z)}}{z^2(q^2z^2-4m^2(z-1))},
\end{gather}
where
\begin{gather*}
P_3(z)=q^2z^2(z-2)^2+4m^2(z-1)^2\big(4+z^2+z^3+z^4+z^5\big).
\end{gather*}

It is easy to see that the function $P_3(z)$ under the radical in (\ref{sp13}), (\ref{sp14}) is strictly positive, so $u_3(z)$ is smooth for all $z>0$. As a consequence, the embedding function (\ref{sp11}) describes the surface which is smooth for all $r>0$. As for two embeddings given above, the surface points which corresponds to $r\to0$ tend to  inf\/inity.

Thus the embedding (\ref{sp11}) is also global (i.e.\ smooth for all $r>0$). We call it a ``cubic embedding''.

\subsection{Area covered by embeddings}\label{p34}

For all the three embeddings given  above the smoothness of the corresponding surfaces is evident when the parameter $t'$ is chosen as a coordinate on the surface. When one turns to the coordina\-te~$t$ (in this case the metric corresponds to the interval~(\ref{1})) a coordinate singularity appears, since the functions $h_{1,2,3}(z)$ which connect~$t'$ and~$t$ (see (\ref{sp2}), (\ref{sp7}), (\ref{sp12})) have logarithmic singularities on the horizons $r=r_{\pm}$ (since the denominators of the integrals (\ref{sp4}), (\ref{sp9}), (\ref{sp14}) are equal to $g_{00}$ up to a factor). It leads to the fact that at $r\to r_++0$ and $t\to +\infty$ the coordinate $t'$ has a f\/inite limit, so on the surface there exists a trajectory that describes the transition of the particle through the exterior horizon of a black hole. If  $r\to r_++0$ and $t\to -\infty$, then $t'\to -\infty$. It means that if one considers the motion in the past along the trajectory corresponding to the emission of a particle from a white hole, for the given embedding the surface point which corresponds to the external horizon tend to inf\/inity. Note that the same situation takes place for three of six Schwarzschild metric embeddings given in~\cite{statja27}.

One can show that in all the three types of embeddings the coordinate $t'$ dif\/fers from the retarded Eddington--Finkelstein time only by a smooth function of $r$, so these embeddings cover not all regions which correspond to a maximal analytical extension of the RN metric, but only those which are described by the Eddington--Finkelstein coordinates, see, for example,~\cite{griffiths}. The covered region is shown on the corresponding Penrose diagram on the f\/igure.

\begin{figure}[h]
\centering
\includegraphics[width=16em]{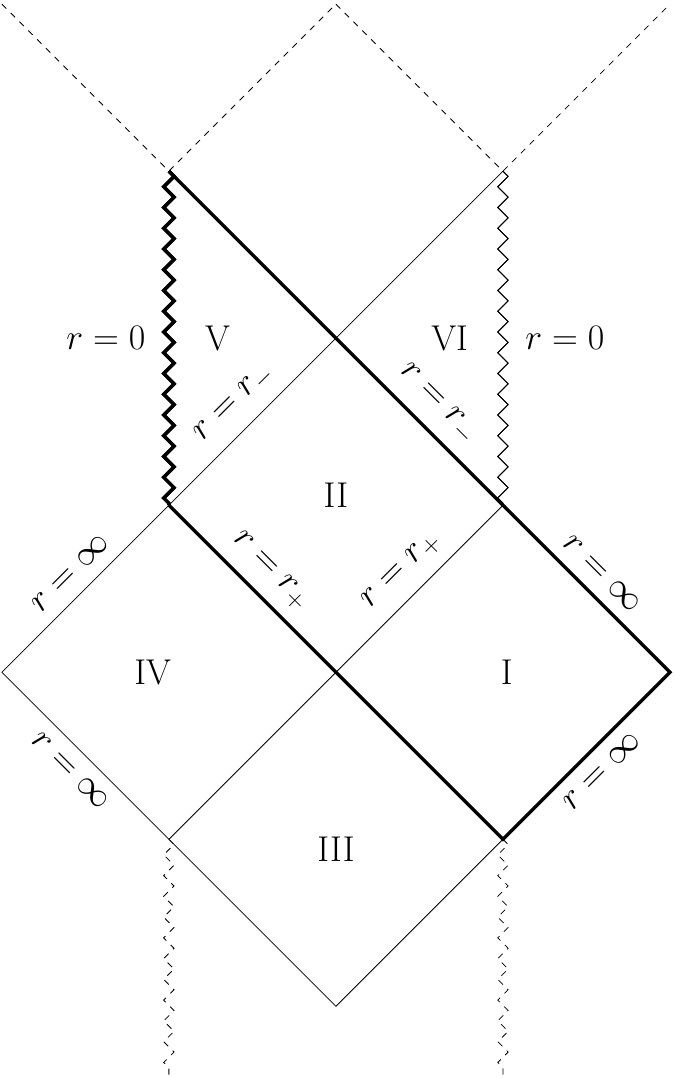}
\caption{Penrose diagram for a maximal analytical extension of the RN metric.
I, IV~-- two regions with $r>r_+$;
II~-- region with $r_-<r<r_+$ corresponding to a black hole;
III~-- region with $r_-<r<r_+$ corresponding to a white hole;
V, VI~-- two regions with $r<r_-$.
A part covered  by the embeddings~(\ref{2}), (\ref{sp6}), (\ref{sp11}) is bordered by the thick line.
}
\end{figure}

\subsection*{Acknowledgements}
This work was partially supported by the Saint Petersburg State University grant\\
11.38.660.2013.



\end{document}